\documentclass[preprintnumbers,10pt,nofootinbib]{revtex4}

\usepackage{amsmath,latexsym,amssymb,amsfonts}
\usepackage{graphicx}
\usepackage{bm}
\usepackage{epstopdf} 

\usepackage{xcolor}

\begin{document}
\title{\textbf{Causal structures and dynamics of black-hole-like solutions in string theory}}

\author{
\textsc{Subeom Kang$^{a}$}\footnote{{\tt ksb527{}@{}kaist.ac.kr }} and
\textsc{Dong-han Yeom$^{b,c,d,e}$}\footnote{{\tt innocent.yeom{}@{}gmail.com}}
}

\affiliation{
$^{a}$Department of Physics, KAIST, Daejeon 34141, Republic of Korea\\
$^{b}$Asia Pacific Center for Theoretical Physics, Pohang 37673, Republic of Korea\\
$^{c}$Department of Physics, POSTECH, Pohang 37673, Republic of Korea\\
$^{d}$Department of Physics Education, Pusan National University, Busan 46241, Republic of Korea\\
$^{e}$Research Center for Dielectric and Advanced Matter Physics, Pusan National University, Busan 46241, Republic of Korea
}

\begin{abstract}
We investigate spherically symmetric solutions in string theory. Such solutions depend on three parameters, one of which corresponds to the asymptotic mass while the other two are the dilaton and two-form field amplitudes, respectively. If the two-form field amplitude is non-vanishing, then this solution represents a trajectory of a singular and null hypersurface. If the dilaton and two-form field amplitudes are non-vanishing but very close to zero, then the solution is asymptotically the same as the Schwarzschild solution, while only the near horizon geometry will be radically changed. If the dilaton field diverges toward the weak coupling regime, this demonstrates a firewall-like solution. If the dilaton field diverges toward the strong coupling limit, then as we consider quantum effects, this spacetime will emit too strong Hawking radiation to preserve semi-classical spacetime. However, if one considers a junction between the solution and the flat spacetime interior, this can allow a stable star-like solution with reasonable semi-classical properties. We discuss possible implications of these causal structures and connections with the information loss problem.
\end{abstract}

\maketitle

\newpage

\tableofcontents

\section{Introduction}

The information loss problem is one of the key unresolved problems in the quantum theory of gravity \cite{Hawking:1974sw,Hawking:1976ra}. There have been various approaches to resolving the paradox. We may classify the possible ideas as follows.
\begin{itemize}
\item[--] 1. \textit{Information loss}. After black hole evaporation, unitarity is violated \cite{Hawking:1976ra}. However, this idea is not consistent with the AdS/CFT correspondence \cite{Maldacena:1997re}; also, there exists a debate as to whether or not the violation of unitarity causes a serious problem \cite{Banks:1983by}. 
\item[--] 2. \textit{Hawking radiation carries information}. If one assumes that general relativity and semi-classical quantum field theory are still valid \cite{Susskind:1993if}, then there appears an inconsistency \cite{Yeom:2008qw}. Hence, if one still assumes that Hawking radiation carries information, this means that the general relativity must be severely modified around the horizon scale, or the so-called firewall should appear \cite{Almheiri:2012rt}. Such a violation of general relativity can be naked and there is no principle to control the firewall \cite{Hwang:2012nn}.
\item[--] 3. \textit{Information is retained by remnants.} This covers a wide range of possibilities, including Planck scale remnants, the baby universe scenario, and so on \cite{Chen:2014jwq}. This idea has several problems with entropy \cite{Hwang:2016otg} and there is no argument to show that the remnant picture can be applied to very generic circumstances.
\item[--] 4. \textit{Effective loss of information.} It is possible to suggest that information is not carried by Hawking radiation nor stored in remnants, but unitarity is still preserved when we consider the entire wave function of the universe \cite{Hawking:2005kf}. This idea requires tunneling from gravitational collapse to a trivial geometry \cite{Sasaki:2014spa}. This can be applied to generic systems \cite{Chen:2017suz}, but one needs to check whether or not this really preserves unitarity.
\end{itemize}
The most preferred idea among string theorists is perhaps the second one. There have been proposals to overcome the violation of general relativity \cite{Maldacena:2013xja}, but these have also faced criticism \cite{Chen:2016nvj}. Unless there exists a way to explain the physical origin of the firewall constructively, the discussion about the firewall remains speculative.

In this context, we propose a model inspired by string theory. We consider a model that includes gravity, the dilaton field, and the Kalb--Ramond field \cite{Burgess:1994kq}. This is motivated not only by traditional string theory but also the recent discussions in the context of double field theory \cite{Siegel:1993fxa}. It was already known that there exists a trivial Schwarzschild solution. One can also construct a non-trivial dilaton and Kalb--Ramond field combination, but the solution becomes singular. This is the reason why many people did not consider this solution seriously for several decades. On the other hand, if we consider double field theory, the singular solution can be interpreted as a vacuum solution of the fundamental action \cite{Ko:2016dxa}. In addition to this, we will revisit the solution and check whether it can work as a kind of firewall. If this is the case, then it can be a very natural way to construct a firewall within the framework of string theory\footnote{It is interesting to compare this idea with the fuzzball picture which is famous in the string theory community \cite{Mathur:2005a}. According to the picture, a black hole is a superposition of various fuzzball solutions, where each solution has no horizon but has a regular membrane. On the other hand, if we superpose all fuzzball states, it is conjectured that the averaged geometry looks like a semi-classical black hole. In this sense, our solution is quite different from the fuzzball conjecture, because our solution is singular near the center and this is no more a superposition of other solutions. This is the reason why we would like to compare this solution with not the fuzzball picture but the firewall conjecture.}.

This paper is organized as follows. In Sec.~\ref{sec:rev}, we revisit the non-trivial solution with the dilaton and the Kalb--Ramond field. In Sec.~\ref{sec:cau}, we discuss details on the causal structures of the solution. We will find that there are two branches of solutions, one corresponding to the strong coupling limit and the other corresponding to the weak coupling limit. In Sec.~\ref{sec:sta}, we discuss more on the former case: by considering a star-like interior, one interprets this as a regular star-like solution without a singularity. In Sec.~\ref{sec:app}, we discuss more on the latter case: in the weak coupling limit, this solution naturally explains the firewall and there may be no problem of information loss. Finally, in Sec.~\ref{sec:con}, we discuss several possible criticisms and possible future work.

\section{\label{sec:rev}Review of the solution}

In this section, we comprehensively review the non-trivial solution which includes the dilaton field and the Kalb--Ramond field. We investigate stringy solutions of the model \cite{Burgess:1994kq}:
\begin{eqnarray}
S = \int dx^{4} \sqrt{-g} e^{-2\phi} \left( \mathcal{R} + 4 \partial_{\mu} \phi \partial^{\mu} \phi - \frac{1}{12} H_{\mu\nu\rho} H^{\mu\nu\rho} \right),
\end{eqnarray}
where $\phi$ is the dilaton field, and $H = dB$ is the field strength tensor of the Kalb--Ramond field $B_{\mu\nu}$.

We specificially focus on the following class of static solutions \cite{Burgess:1994kq,Ko:2016dxa}:
\begin{eqnarray}\label{eq:metform}
ds^{2} = e^{2\phi(r)} \left( -A(r) dt^{2} + \frac{1}{A(r)} dr^{2} + \frac{C(r)}{A(r)}d\Omega^{2} \right),
\end{eqnarray}
where
\begin{eqnarray}
e^{2\phi} &=& \gamma_{+} \left( \frac{r - \alpha}{r + \beta} \right)^{\frac{b}{\sqrt{a^{2} + b^{2}}}} + \gamma_{-} \left( \frac{r - \alpha}{r + \beta} \right)^{-\frac{b}{\sqrt{a^{2} + b^{2}}}},\\
A(r) &=& \left( \frac{r-\alpha}{r+\beta} \right)^{\frac{a}{\sqrt{a^{2}+b^{2}}}},\\
C(r) &=& (r-\alpha) (r+\beta),\\
\alpha &\equiv& \frac{a}{a+b} \sqrt{a^{2}+b^{2}},\\
\beta &\equiv& \frac{b}{a+b} \sqrt{a^{2}+b^{2}},\\
\gamma_{\pm} &\equiv& \frac{1}{2} \left( 1 \pm \sqrt{1 - \frac{h^{2}}{b^{2}}} \right).
\end{eqnarray}
We can transform this metric as
\begin{eqnarray}
ds^{2} = - F(R) dt^{2} + G(R) dR^{2} + R^{2} d\Omega^{2},
\end{eqnarray}
where
\begin{eqnarray}
R(r) &=& \sqrt{e^{2\phi(r)} \frac{C(r)}{A(r)}},\\
F(R(r)) &=& e^{2\phi(r)} A(r),\\
G(R(r)) &=& \frac{e^{2\phi(r)}}{A(r) (dR/dr)^{2}}.
\end{eqnarray}
Eventually, one will be able to determine $F$ and $G$ as functions of $R$. Note that the physical meanings of $a$, $b$, and $h$ are given by \cite{Ko:2016dxa}. Here, $h$ represents the field value of the Kalb-Ramond field, where this two-form field becomes $B_{(2)} = h \cos\theta dt \wedge d\varphi$. Regarding $a$ and $b$, it is not easy to assign the direct physical meanings, but if $a \gg 1$ and $b, h \ll 1$, then $a$ can be interpreted as the ADM mass, where we will mainly consider this limit. More detailed interpretation of the conservative quantities is discussed in \cite{Ko:2016dxa}.  

\begin{figure}
\begin{center}
\includegraphics[scale=0.8]{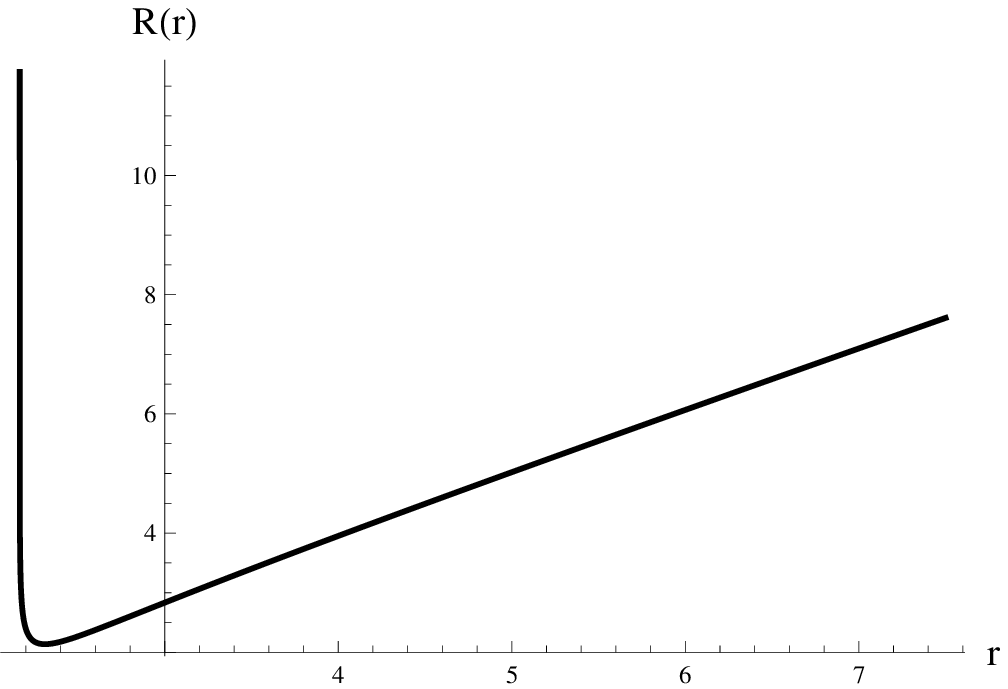}
\includegraphics[scale=0.8]{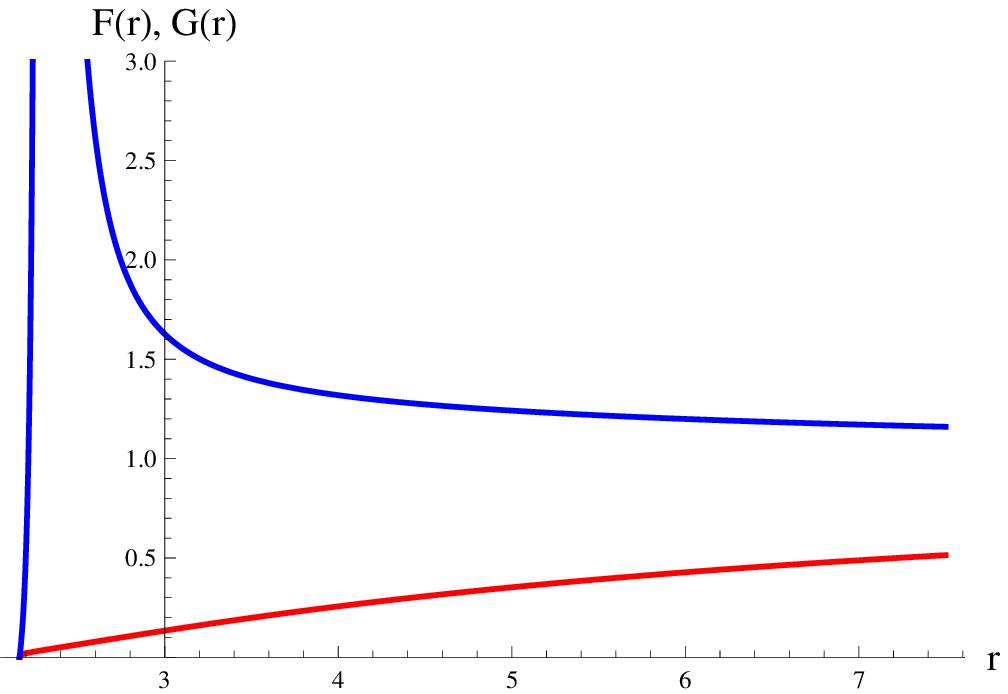}
\caption{\label{fig:JF}Typical behaviors of $R$ (left), $F$ (right, red), and $G$ (right, blue), where we have chosen $a=3$, $b=2$, and $h=1$. Even though the dilaton field is singular at the center, as long as $a > b$, the center is null. One can see that $R$ is bounced to infinity and hence the center is a kind of infinity.}
\includegraphics[scale=0.8]{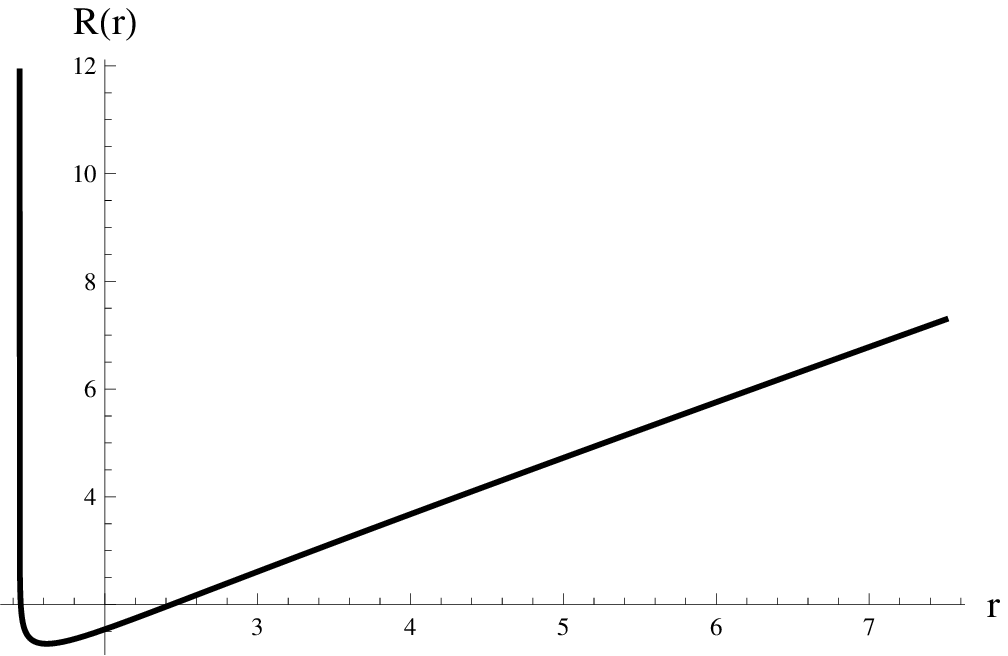}
\includegraphics[scale=0.8]{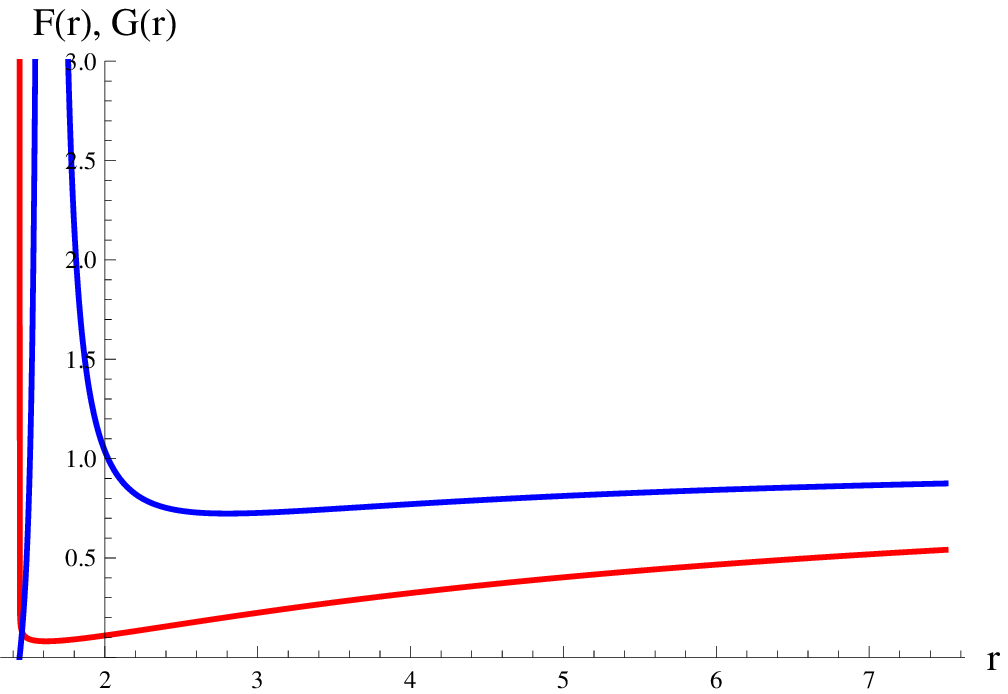}
\caption{\label{fig:JF2}Typical behaviors of $R$ (left), $F$ (right, red), and $G$ (right, blue), where we have chosen $a=2$, $b=3$, and $h=1$. For $b \geq a$, the center is timelike since $F$ is positive.}
\includegraphics[scale=0.8]{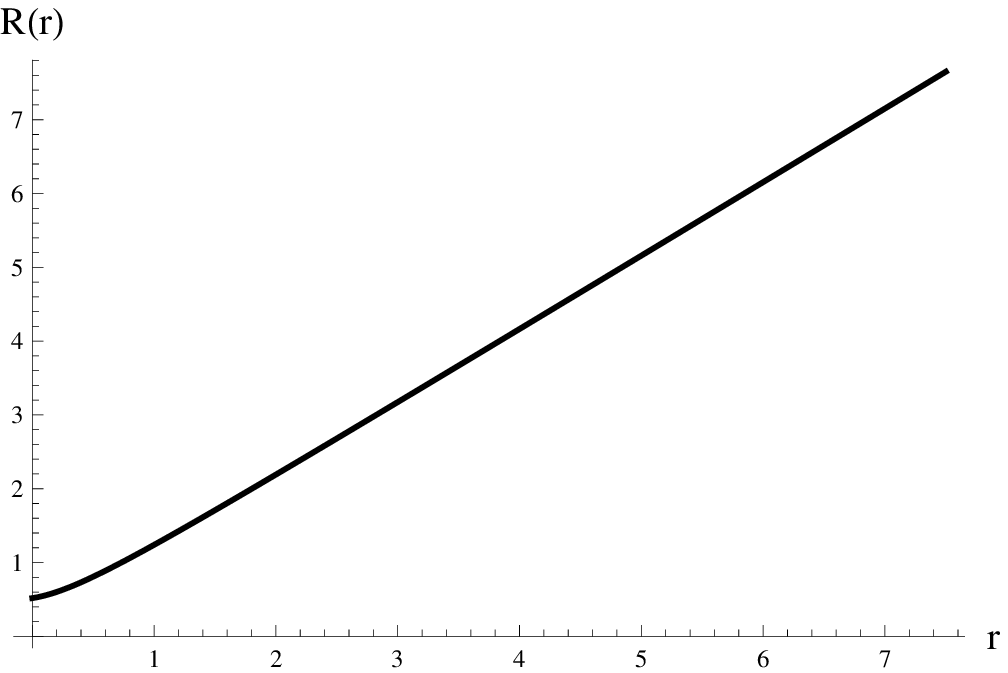}
\includegraphics[scale=0.8]{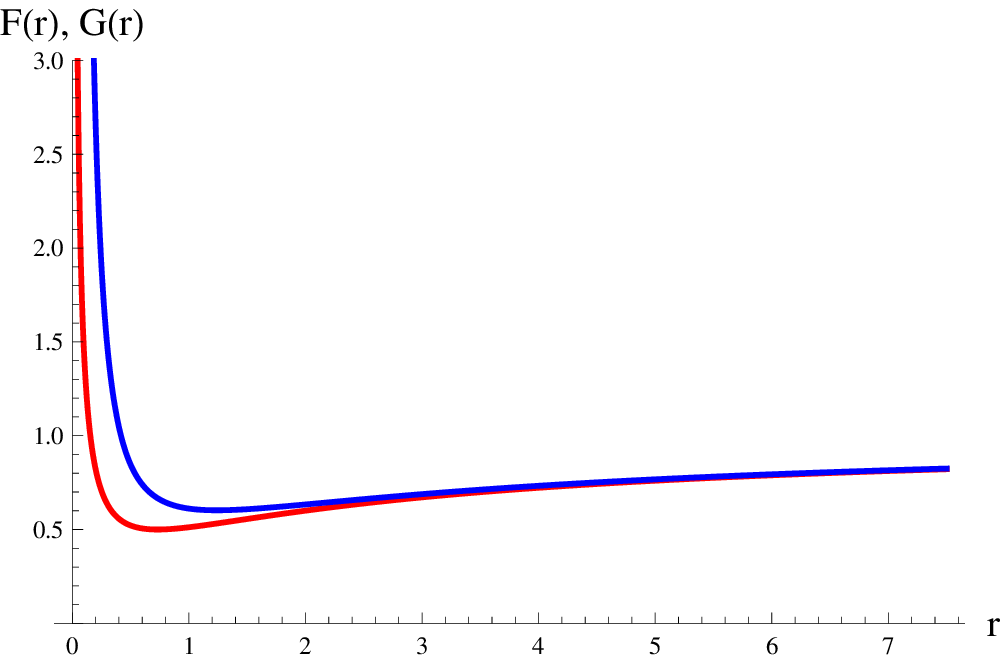}
\caption{\label{fig:JF3}Typical behaviors of $R$ (left), $F$ (right, red), and $G$ (right, blue), where we have chosen $a=0$, $b=2$, and $h=1$. At finite $R$, $F$ and $G$ diverge and terminate, while the surface is timelike.}
\end{center}
\end{figure}

\section{\label{sec:cau}Causal structures of singular null-shell solutions}

\subsection{Jordan frame}

From the solution, there are some simple things that we can check. As $r$ approaches to $\alpha$, $C$ approaches to zero but $R^{2} \propto (r - \alpha)^{1 - (a+b)/\sqrt{a^2 + b^2}}$. This exponent is negative unless $a$ or $b$ is zero; if $a$ or $b$ is zero, then the exponent is zero. Therefore, if both $a$ and $b$ are positive, then as $r$ approaches to $r_{\mathrm{min}} = \alpha$, the areal radius $R$ tends to infinity.

Note that at the point $r_{\mathrm{min}}$, $G$ is always zero, while $F \propto (r-\alpha)^{(a-b)/\sqrt{a^2 + b^2}}$. Therefore, if $a > b > 0$, then $F$ goes to zero at $r = r_{\mathrm{min}}$ and hence the surface is null. If $b > a > 0$, then $F$ becomes positive and the $r = r_{\mathrm{min}}$ surface becomes timelike. Between the two hypersurfaces $r = \infty$ (asymptotic infinity) and $r = r_{\mathrm{min}}$, there must exist a surface that satisfies $R' = 0$, i.e., a bottleneck. Note that the $R' = 0$ surface is always timelike.

Finally, it is easy to check that the Ricci scalar should be divergent at $r_{\mathrm{min}}$:
\begin{eqnarray}
\mathcal{R} = - \frac{F''}{FG} + \frac{F'^{2}}{2F^{2}G} + \frac{F'G'}{2FG^{2}} - \frac{2F'}{FGR} + \frac{2G'}{G^{2}R} - \frac{2}{GR^{2}} + \frac{2}{R^{2}}.
\end{eqnarray}

\begin{figure}
\begin{center}
\includegraphics[scale=0.8]{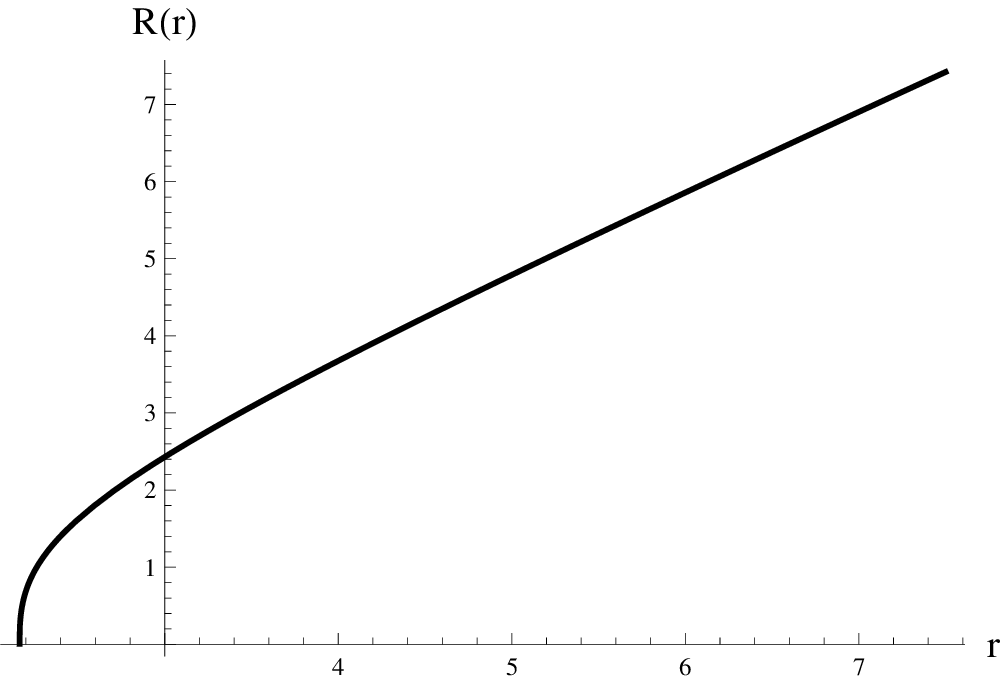}
\includegraphics[scale=0.8]{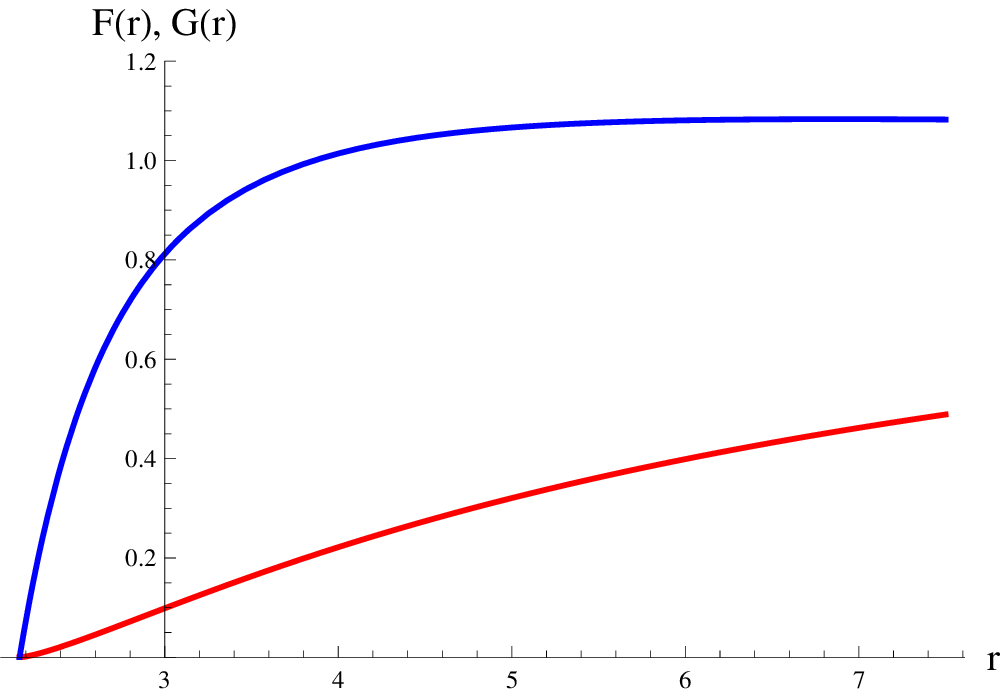}
\caption{\label{fig:JF4}Typical behaviors of $R$ (left), $F$ (right, red), and $G$ (right, blue), where we have chosen $a=3$, $b=2$, and $h=0$. At $R = 0$, $F$ and $G$ both approach to zero and hence the surface is null.}
\end{center}
\end{figure}

\begin{figure}
\begin{center}
\includegraphics[scale=0.7]{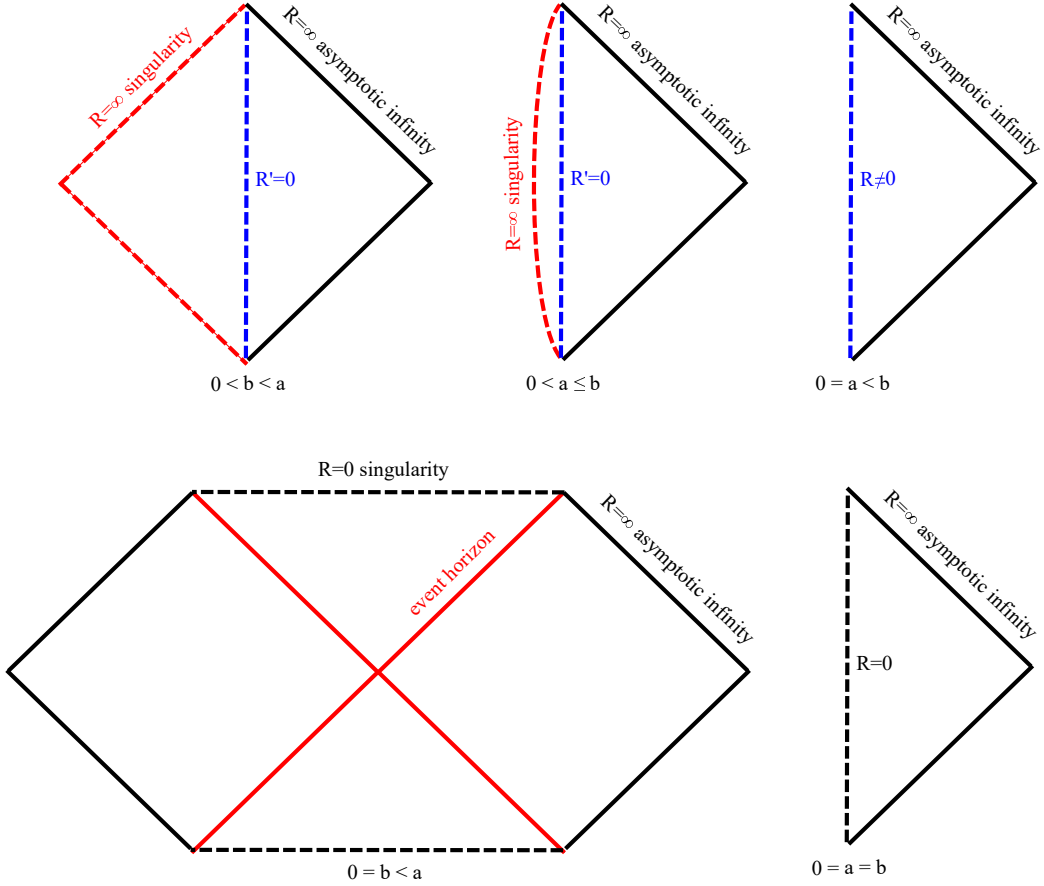}
\caption{\label{fig:ccausal2}Causal structures of all possible parameters ($h > 0$).}
\end{center}
\end{figure}

Then, for $h > 0$, we can categorize five cases and draw their causal structures.
\begin{description}
\item[-- Case 1:] For $0 < b < a$, the bottleneck $R' = 0$ is timelike and the $r = r_{\mathrm{min}}$ singularity is null (see Fig.~\ref{fig:JF}). $G$ and $F$ are positive except at $r_{\mathrm{min}}$, while at the point $r_{\mathrm{min}}$, the surface becomes null.
\item[-- Case 2:] If $0 < a \leq b$, then both $R' = 0$ and $r = r_{\mathrm{min}}$ surfaces are timelike (Fig.~\ref{fig:JF2}).
\item[-- Case 3:] For $0 < b$ and $a = 0$, $R$ becomes singular at non-zero values (Fig.~\ref{fig:JF3}).
\item[-- Case 4:] For $0 < a$ and $b = 0$, the metric is trivial Schwarzschild.
\item[-- Case 5:] For $a = b = 0$, the metric is trivial Minkowski.
\end{description}
However, if $h = 0$, then $R$, $F$, and $G$ all approach to zero and hence it becomes an $R = 0$ null singularity (Fig.~\ref{fig:JF4}). The causal structures are summarized by Fig.~\ref{fig:ccausal2}.

\subsection{Einstein frame}

In the Einstein frame ($g^{\mathrm{E}}_{\mu\nu} = g_{\mu\nu} e^{-2\phi}$), the solution is simplified such that
\begin{eqnarray}
ds_{\mathrm{E}}^{2} = -A(r) dt^{2} + \frac{1}{A(r)} dr^{2} + \frac{C(r)}{A(r)}d\Omega^{2}.
\end{eqnarray}
This can be also modified as follows:
\begin{eqnarray}
ds_{\mathrm{E}}^{2} = - F_{\mathrm{E}}(R_{\mathrm{E}}) dt^{2} + G_{\mathrm{E}}(R_{\mathrm{E}}) dR_{\mathrm{E}}^{2} + R_{\mathrm{E}}^{2} d\Omega^{2},
\end{eqnarray}
where
\begin{eqnarray}
R_{\mathrm{E}}(r) &=& \sqrt{\frac{C(r)}{A(r)}},\\
F_{\mathrm{E}}(R_{\mathrm{E}}(r)) &=& A(r),\\
G_{\mathrm{E}}(R_{\mathrm{E}}(r)) &=& \frac{1}{A(r) (dR_{\mathrm{E}}/dr)^{2}}.
\end{eqnarray}
In the Einstein frame, there is no dependence of the metric on $h$.

\begin{figure}
\begin{center}
\includegraphics[scale=0.8]{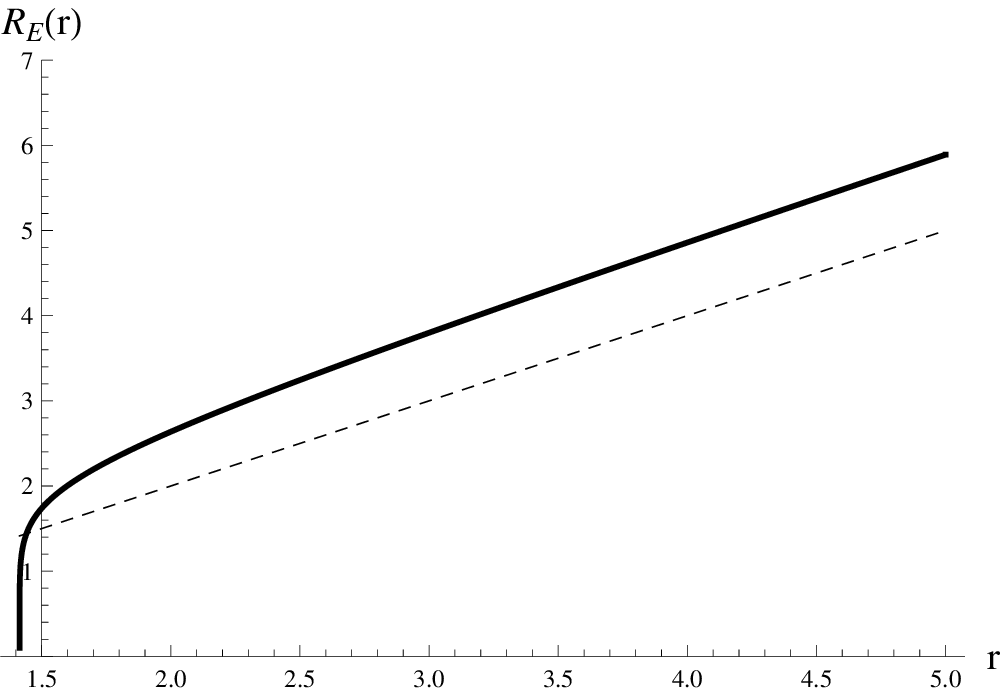}
\includegraphics[scale=0.8]{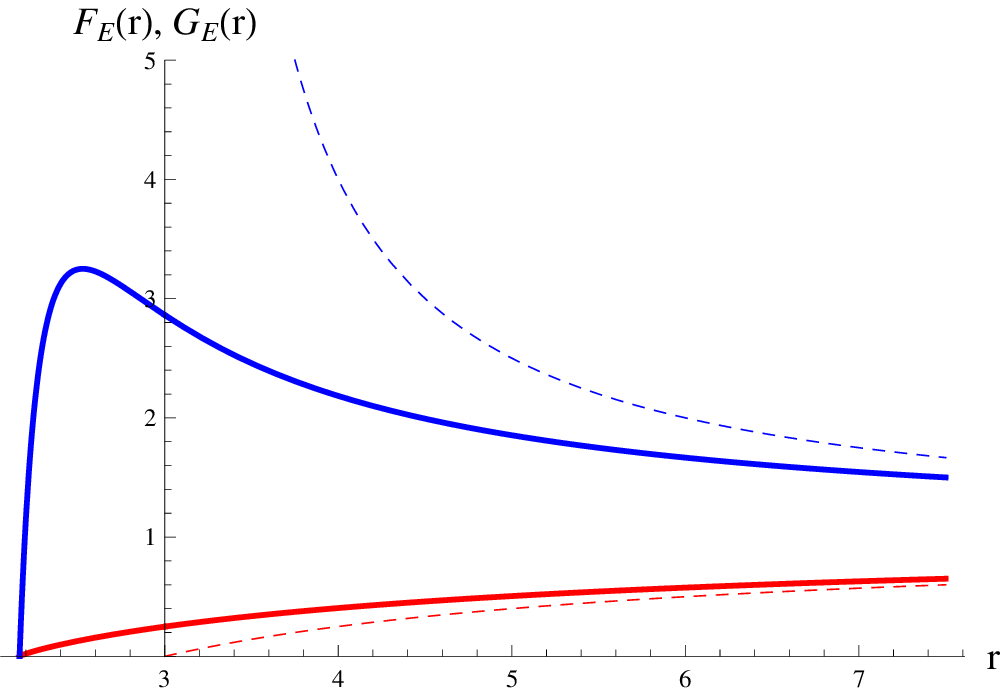}
\caption{\label{fig:EF}Typical behaviors of $R_{\mathrm{E}}$ (left), $F_{\mathrm{E}}$ (right, red), and $G_{\mathrm{E}}$ (right, blue), where we have chosen $a=3$ and $b=2$. Dashed curves correspond to $R_{\mathrm{E}}$ (black), $F_{\mathrm{E}}$ (red), and $G_{\mathrm{E}}$ (blue), respectively, for $a=3$ and $b=0$ (i.e. the Schwarzschild solution). As $r$ goes to infinity, the solution approaches the Schwarzschild solution. On the other hand, as $r$ goes to zero, $R_{\mathrm{E}}$ rapidly drops to zero and both of $F_{\mathrm{E}}$ and $G_{\mathrm{E}}$ become zero. Therefore, $R_{\mathrm{E}} = 0$ is a null surface as long as $a, b > 0$.}
\includegraphics[scale=0.8]{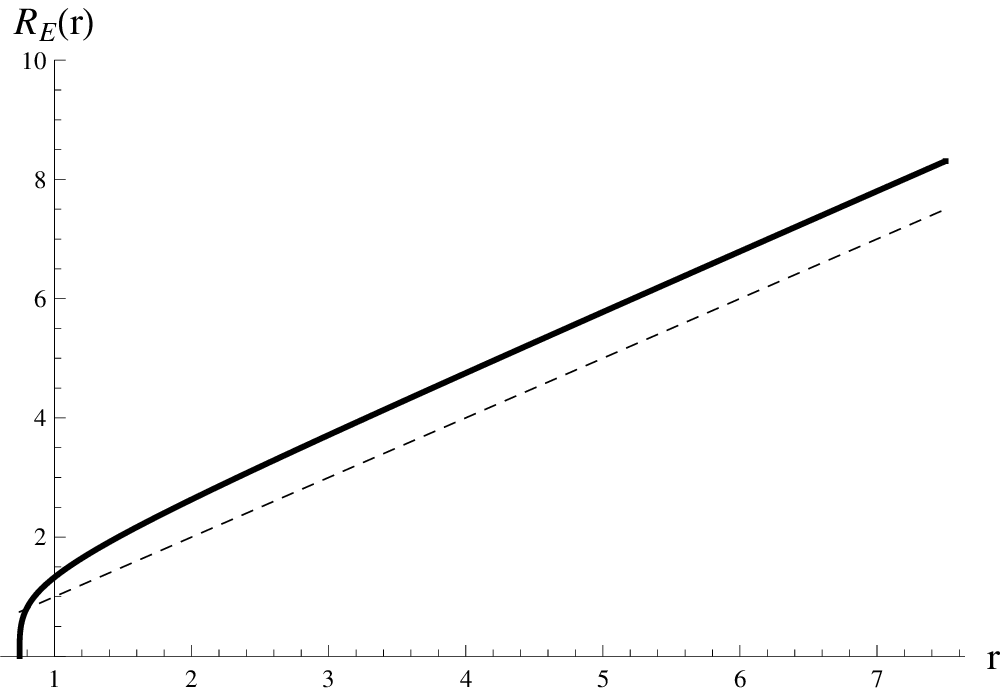}
\includegraphics[scale=0.8]{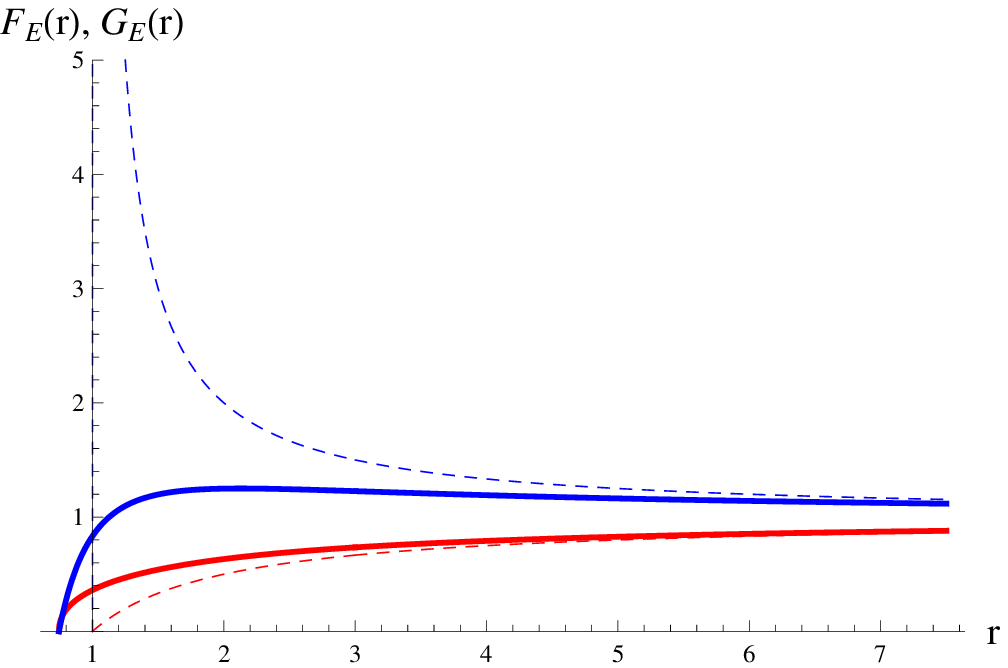}
\caption{\label{fig:EF2}Typical behaviors of $R_{\mathrm{E}}$ (left), $F_{\mathrm{E}}$ (right, red), and $G_{\mathrm{E}}$ (right, blue), where we have chosen $a=1$ and $b=2$. Dashed curves correspond to $R_{\mathrm{E}}$ (black), $F_{\mathrm{E}}$ (red), and $G_{\mathrm{E}}$ (blue), respectively, for $a=1$ and $b=0$ (i.e. the Schwarzschild solution).}
\end{center}
\end{figure}

In the Einstein frame, we can observe that $R_{\mathrm{E}} = 0$ if $r = \alpha$, since
\begin{eqnarray}
R_{\mathrm{E}}^{2} = \left( r - \alpha \right)^{1-\frac{a}{\sqrt{a^{2}+b^{2}}}} \left( r + \beta \right)^{1+\frac{a}{\sqrt{a^{2}+b^{2}}}},
\end{eqnarray}
except for $b = 0$. At the same time, $F_{\mathrm{E}}(R_{\mathrm{E}}=0) = G_{\mathrm{E}}(R_{\mathrm{E}}=0) = 0$. These observations imply that this solution generically (if $b$ is not zero) has a null singularity at the center $R_{\mathrm{E}} = 0$ (Fig.~\ref{fig:EF}). However, in the Jordan frame, if $b > a > 0$, then $r = \alpha$ is timelike, while the surface is null in the Einstein frame (see Fig.~\ref{fig:EF2}). This inequivalence is due to the singular conformal transformation.

If $a=0$ but $b$ remains finite, then the metric reduces to the form
\begin{eqnarray}
ds_{\mathrm{E}}^{2} &=& -dt^{2} + dr^{2} + r(r+b)d\Omega^{2}\\
&=& -dt^{2} + \frac{1}{1+b^{2}/4R_{\mathrm{E}}^{2}}dR_{\mathrm{E}}^{2} + R_{\mathrm{E}}^{2}d\Omega^{2},
\end{eqnarray}
which has a timelike singularity at the center.

\subsection{Two branches of solutions}

Even though there is no horizon, since the asymptotic structure is the same as Schwarzschild, one can expect Hawking radiation and emission of particles \cite{Davies:1976ei}. This will cause $a$ to decrease.

If $h > 0$, then near the singular null surface, $\phi$ goes to $+\infty$. This means that the gravitational coupling will become stronger and stronger, and hence this solution will rapidly evaporate and disappear. Hence, this means that the evaporation time of the solution is faster than that of the formation time: the solution will disappear before the singularity is formed.

One exception is to consider $h = 0$. Then near the horizon, $\phi$ approaches to $-\infty$ and hence the solution approaches the weak coupling limit. Then we can give a reasonable physical interpretation of the semi-classical evaporation of the solution. If $b$ is small enough, then the geometry is almost the same as Schwarzschild, while the geometry is suddenly modified around the horizon scale $r \lesssim 2M \sim a$. Following this, it quickly becomes a singularity. Hence, effectively, there is no interior of the black hole.

We discuss two possibilities separately in the following sections. In Sec.~\ref{sec:sta}, we discuss the strong coupling case. In order to avoid such a strong coupling limit, we consider a junction with a stable star-like interior. As the black hole evaporates, the mass will slowly decrease and eventually the star-like interior will form a black hole. It will then quickly disappear due to the strong evaporation. In Sec.~\ref{sec:app}, we discuss the weak coupling case. There then appears a stable but singular null surface which can be interpreted as a natural realization of a firewall.

\section{\label{sec:sta}Strong coupling limit: star-like interior solutions}

\subsection{Thin-shell approximation}

In order to extend to the $h > 0$ limit, we consider a junction with a thin-shell \cite{Israel:1966rt} such that the outside is the stringy solution while the inside is Minkowski without a singularity nor a strong coupling limit. Hence, for the outside we assume $a > b$ and $h > 0$, while for the inside we take $a = b = h = 0$.

As we are considering the thin-shell approximation, the metric ansatz is
\begin{eqnarray}
ds_{\pm}^{2} = - e^{2\Phi_{\pm}(R)}f_{\pm}(R) dt^{2} + \frac{1}{f_{\pm}(R)} dR^{2} + R^{2} d\Omega^{2},
\end{eqnarray}
where $\pm$ means outside and inside the shell, respectively. The induced metric on the shell is assumed to be
\begin{eqnarray}
ds_{\mathrm{shell}}^{2} = - d\tau^{2} + R(\tau)^{2} d\Omega^{2}.
\end{eqnarray}
Since $R$ is a function of $r$, we assume here that $R(r(\tau))$.

The inside geometry is assumed to be Minkowski:
\begin{eqnarray}
ds_{-}^{2} = - dt^{2} + dR^{2} + R^{2} d\Omega^{2}.
\end{eqnarray}
For the outside, we can transform the metric as
\begin{eqnarray}
ds^{2} = - e^{2\Phi_{+}(R)}f_{+}(R) dt^{2} + \frac{1}{f_{+}(R)} dR^{2} + R^{2} d\Omega^{2},
\end{eqnarray}
where
\begin{eqnarray}
R(r) &=& \sqrt{e^{2\phi(r)} \frac{C(r)}{A(r)}},\\
\Phi_{+}(R(r)) &=& \log \frac{e^{2\phi(r)}}{(dR/dr)},\\
f_{+}(R(r)) &=& \frac{A(r) (dR/dr)^{2}}{e^{2\phi(r)}}.
\end{eqnarray}
Regarding the field values, the solution becomes
\begin{eqnarray}
e^{2 \phi_{-}(r)} &=& 1,\\
h_{-}(r) &=& 0
\end{eqnarray}
for inside the shell and
\begin{eqnarray}
e^{2 \phi_{+}(r)} &=& \gamma_{+} \left( \frac{r - \alpha}{r + \beta} \right)^{\frac{b}{\sqrt{a^{2} + b^{2}}}} + \gamma_{-} \left( \frac{r - \alpha}{r + \beta} \right)^{-\frac{b}{\sqrt{a^{2} + b^{2}}}},\\
h_{+}(r) &=& h
\end{eqnarray}
for outside the shell, where $B_{(2)}^{\pm} = h_{\pm}(r) \cos\theta dt \wedge d\varphi$. These field values should be continuously connected at the thin transition region (thin-shell). These contributions will give a non-trivial tension and the pressure to the shell\footnote{In the original double field theory point of view, the dilaton and the Kalb-Ramond fields are parts of stringy gravitational fields. However, one can effectively regard them as usual matter fields and they can couple to extra matter fields. If there is no coupling between the dilaton or the Kalb-Ramond fields and the extra matter fields of the thin-shell, then the only allowed solution would be the Schwarzschild solution due to the Birkhoff theorem. Therefore, it is important to notice that there should be a coupling between the extra fields of the matter shell and the dilaton or the Kalb-Ramond field.}.

The Israel junction equation \cite{Israel:1966rt,Garcia:2011aa} is
\begin{eqnarray}
\epsilon_{-} \sqrt{f_{-}(R) + \dot{R}^{2}} - \epsilon_{+} \sqrt{f_{+}(R) + \dot{R}^{2}} = 4\pi \sigma(R) R,
\end{eqnarray}
where $\sigma(R)$ is the tension and $\epsilon_{\pm}$ denotes the outward normal direction: from the causal structures of the static solution in our setup, there is no region with $\epsilon_{\pm} < 0$ and we further assume $\epsilon_{\pm} = +1$. The junction equation is simplified to
\begin{eqnarray}
\left(\frac{dR}{d\tau}\right)^{2} + V(R) = 0,
\end{eqnarray}
where
\begin{eqnarray}
V(R) = f_{+}(R)- \frac{\left(f_{-}(R)-f_{+}(R)-16 \pi^{2} \sigma^{2}(R) R^{2}\right)^{2}}{64\pi^{2}\sigma^{2}(R) R^{2}}.
\end{eqnarray}
Since there exists an implicit dependence on $r$, we obtain
\begin{eqnarray}
\left(\frac{dr}{d\tau}\right)^{2} \left(\frac{dR}{dr}\right)^{2} + V(r) = 0.
\end{eqnarray}
The only allowed solution is $V(r) \leq 0$.

The corresponding pressure that satisfies the conservation equation is \cite{Garcia:2011aa}
\begin{eqnarray}
\lambda(R) = - \frac{R}{2} \left( \sigma'(R) - \Xi(R) \right) -\sigma(R),
\end{eqnarray}
where
\begin{eqnarray}
\Xi(R) \equiv \frac{1}{4\pi R} \left( \Phi_{+}'(R) \sqrt{f_{+}(R) + \dot{R}^{2}} - \Phi_{-}'(R) \sqrt{f_{-}(R) + \dot{R}^{2}}\right).
\end{eqnarray}
If the interior geometry is Minkowski, then
\begin{eqnarray}
\lambda(R(r)) = \frac{1}{8\pi (dR/dr)} \left( 2 \frac{d\phi}{dr} - \frac{d^{2}R/dr^{2}}{dR/dr} - 4\pi R \frac{d\sigma}{dr} \right) \sqrt{f_{+} - V(r)} - \sigma(r).
\end{eqnarray}
Note that if
\begin{eqnarray}
2 \frac{d\phi}{dr} - \frac{d^{2}R/dr^{2}}{dR/dr} = 4\pi R \frac{d\sigma}{dr}
\end{eqnarray}
is satisfied, then the equation of state of the shell is constant, $w = \lambda/\sigma = -1$, and hence this can be constructed with physically reasonable matter. 
In order to simulate a realistic star interior, it is reasonable that there are various contributions to the tension and the pressure including not only from the dilaton and the Kalb-Ramond fields but also from other kinds of interactions. It must be complicated in general, but if there are no singular behaviors and if the contributions from the dilaton and the Kalb-Raomnd fields are sufficiently small, the dominant contribution for the large $r$ limit can be a constant with the equation of state $-1$ following a scalar field. (The condition for the absence of singular behaviors from the dilaton or the Kalb-Ramond field is discussed in Appendix.) So, for simplicity, in this paper, we fix $w = -1$ and solve the tension according to this equation of state:
\begin{eqnarray}
\sigma(r) = \sigma_{0} + \int_{r_{0}}^{r} \frac{1}{4\pi R(\bar{r})} \left( 2 \frac{d\phi(\bar{r})}{d\bar{r}} - \frac{d^{2}R(\bar{r})/d\bar{r}^{2}}{dR(\bar{r})/d\bar{r}} \right) d\bar{r},
\end{eqnarray}
where $\sigma_{0}$ and $r_{0}$ are constants.

\subsection{Fate of the strong coupling limit}

\begin{figure}
\begin{center}
\includegraphics[scale=0.9]{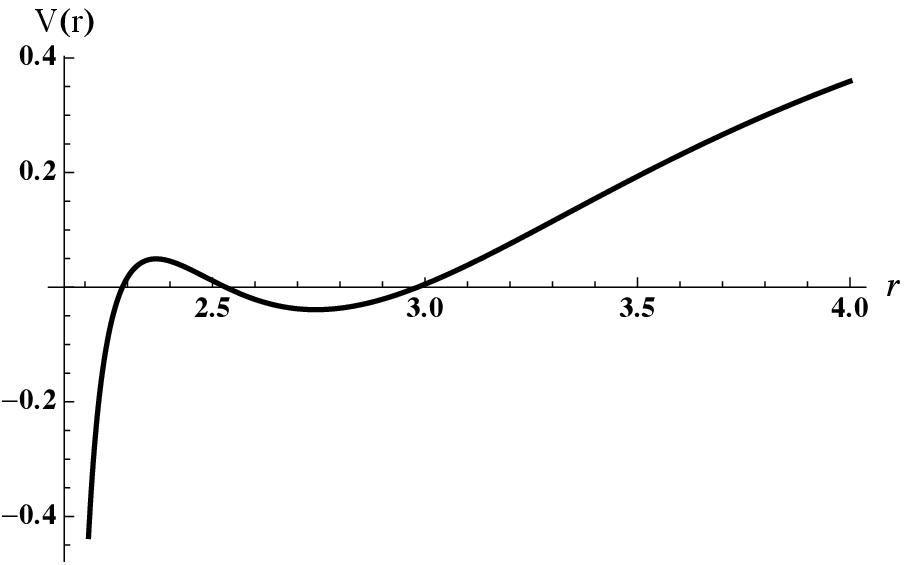}
\includegraphics[scale=0.9]{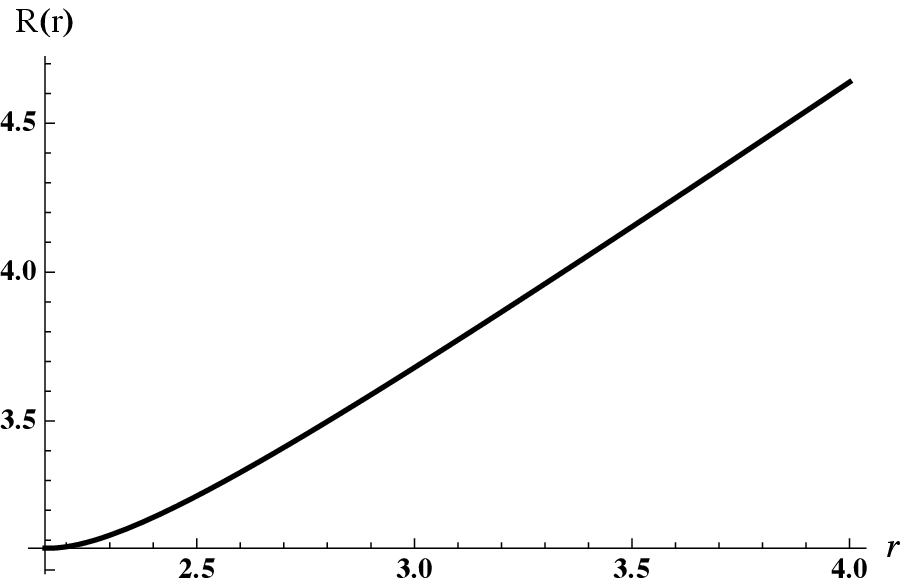}
\includegraphics[scale=0.9]{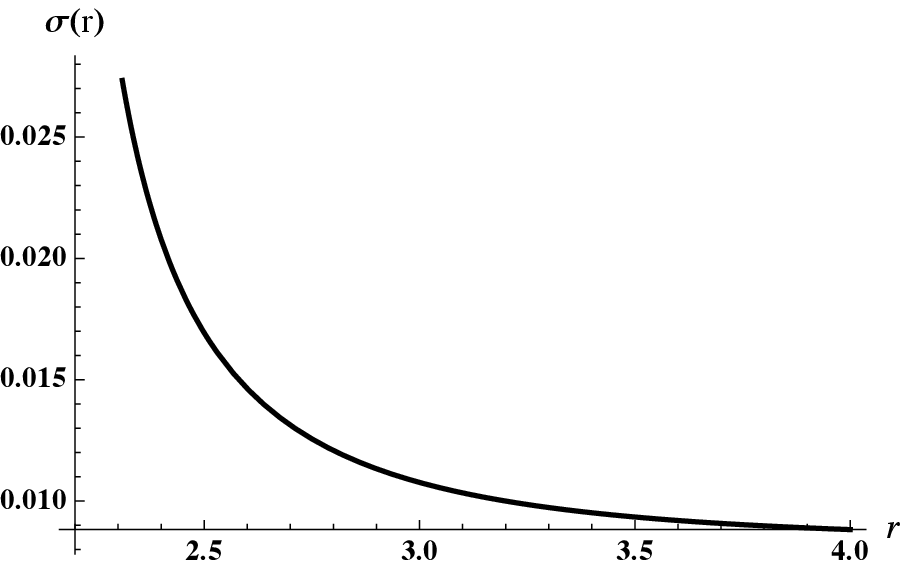}
\includegraphics[scale=0.9]{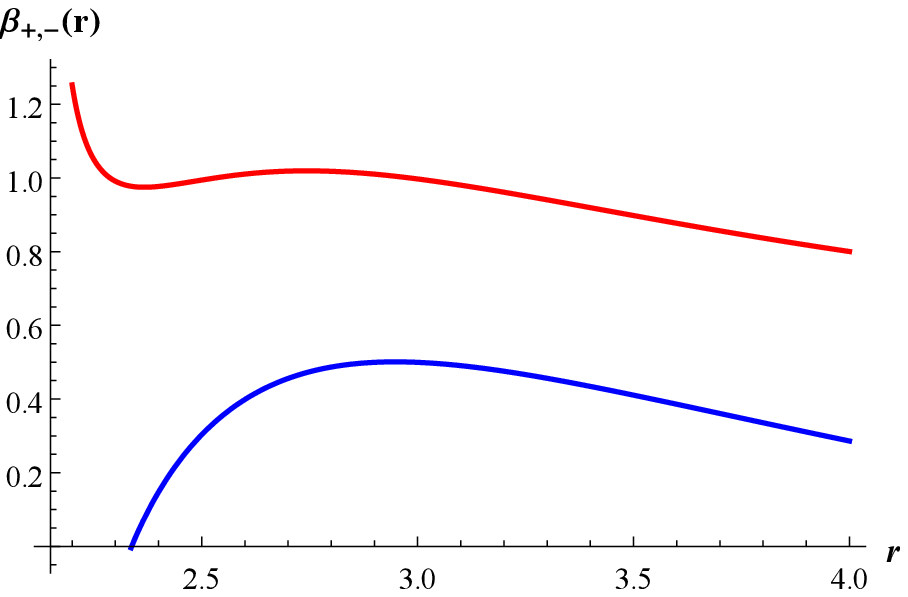}
\caption{\label{fig:example}An example of numerical results (the effective potential $V$, the areal radius $R$, the tension $\sigma$, and the extrinsic curvatures $\beta_{+}$ (red) and $\beta_{-}$ (blue), respectively) for $a=2.5$, $b=1$, $h=0.99$, and $\sigma_{0} = 0.01$.}
\end{center}
\end{figure}

By choosing appropriate parameters, one can find an example that satisfies oscillatory solutions. This is due to the repulsive nature of the near-horizon geometry. By tuning those parameters, one can find a stable stationary solution. Of course, for consistency, we need to check that the signs of the extrinsic curvatures are positive,
\begin{eqnarray}
\beta_{\pm} = \frac{f_{-} - f_{+} \mp 16 \pi^{2} \sigma^{2} R^{2}}{8\pi \sigma R}.
\end{eqnarray}
Fig.~\ref{fig:example} shows such an example.

\begin{figure}
\begin{center}
\includegraphics[scale=0.9]{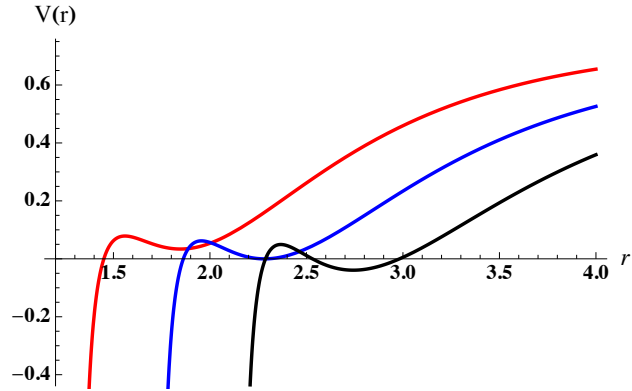}
\caption{\label{fig:moving_V}$V$ as one varies $a$, where $a = 2.5$ (black), $2$ (blue), and $1.5$ (red), respectively.}
\end{center}
\end{figure}

If there exists Hawking radiation, the mass of the black hole $a$ will monotonically decrease. Although it depends on the details of the parameters, one can show an example of the dependences of varying $a$ (Fig.~\ref{fig:moving_V}). This shows that as $a$ decreases, the stable and oscillatory radius decreases and eventually the oscillatory region disappears. There are several ways to interpret this.
\begin{itemize}
\item[--] As $a$ decreases, the oscillatory region disappears. This means that as $a$ becomes smaller and smaller, the thin-shell approximation no longer holds.
\item[--] However, the dependence shows that the radius decreases. Hence, the star-like interior becomes smaller and smaller.
\item[--] Therefore, it is not surprising that as $a$ decreases, eventually the star-like interior will form a black hole. The shell can tunnel to a small shell to collapse; otherwise, as $a$ decreases, the thin-shell approximation breaks and a black hole will be formed.
\end{itemize}

\begin{figure}
\begin{center}
\includegraphics[scale=0.9]{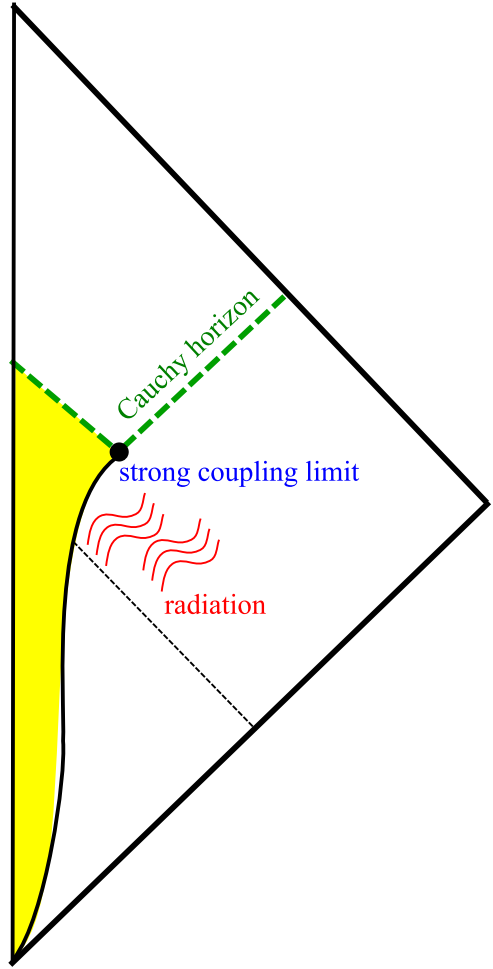}
\caption{\label{fig:ccausal0}Causal structure of stringy regular stars. As the black hole emits radiation, $a$ will monotonically decrease and will approach the strong coupling limit. After that point, the black hole will quickly disappear due to the strong radiation.}
\end{center}
\end{figure}

If $a$ decreases due to the evaporation and the shell approaches the singularity, then the star-like solution will live a sufficiently long time but eventually reach the strong coupling limit. At the critical point, the star-like solution will quickly evaporate. The causal structure is summarized in Fig.~\ref{fig:ccausal0}. At the strong coupling limit, if the spacetime is branched into two pieces, there the asymptotic observer may effectively lose information, but globally information will be preserved, which is very similar to the bubble universe scenario \cite{Hansen:2009kn}. Otherwise, if there is no branch into two pieces, then the spacetime is causally connected and there will be no information loss.

\section{\label{sec:app}Weak coupling limit: alternative to the firewall}

If $h = 0$ and the solution approaches the weak coupling limit, then there is no interior of the black hole and all of spacetime is causally connected (even though there is a singularity). Hence, presumably, there is no information loss. In addition, this solution automatically makes the region inside the apparent horizon (of a Schwarzschild solution) a singularity. This can be an alternative to the firewall.

Then $a$ will decrease. Note that $b$ can be either fixed or decreased to zero. By varying $a$, one can draw the approximate causal structure of an evaporating singular null particle. As $a$ approaches zero, the causal structure will be connected to Minkowski if $b$ also becomes zero.

\begin{figure}
\begin{center}
\includegraphics[scale=0.9]{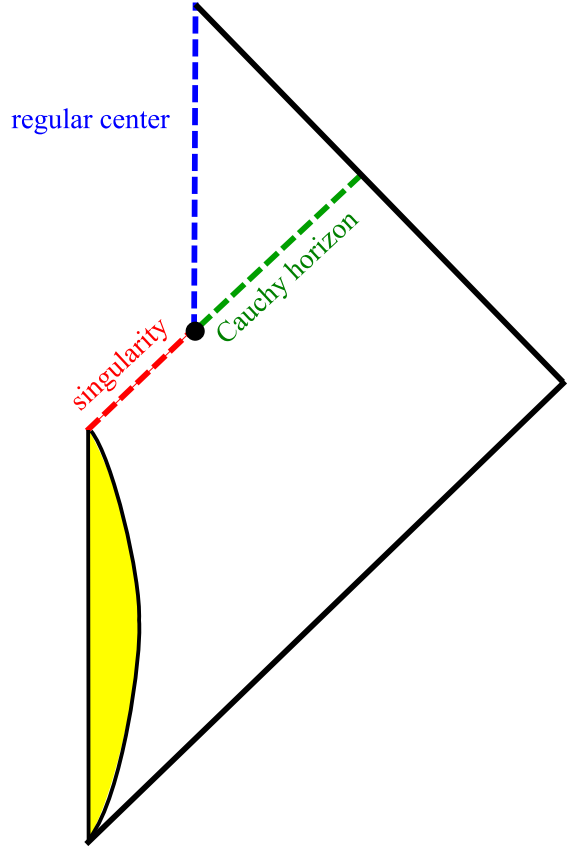}
\caption{\label{fig:ccausal}Causal structure of stringy singular stars. Gravitational collapse (yellow colored region) of a star creates a black hole. If $b$ is turned on at the same time, then a star-like structure is maintained, which will have a null singularity at the center. If the black hole evaporates within a finite time and approaches to $a = b = 0$, then after the evaporation, the center will become timelike again, while the timelike center is regular. If the black hole tends towards a stable object due to the weak coupling limit, then there is no transition to a timelike center and the null singularity will be extended to infinity.}
\end{center}
\end{figure}

To summarize, there are three possibilities:
\begin{itemize}
\item[-- 1.] Both $a$ and $b$ decrease to zero within a finite time: then the black hole will be turned into a Minkowski spacetime.
\item[-- 2.] Only $a$ decreases within a finite time: then the black hole will maintain a null singularity.
\item[-- 3.] If the Hawking temperature decreases to zero, then the black hole can become a stable object maintaining the null singularity.
\end{itemize}
The generic causal structure is summarized in Fig.~\ref{fig:ccausal}.

\section{\label{sec:con}Conclusion}

In this paper, we revisited the singular solutions in string theory. Due to the non-trivial contributions of the dilaton field and the Kalb-Ramond field, the geometry is radically modified to a singularity near the Schwarzschild radius length scale. This does not seem to be a sensible property, but in the context of evaporating black holes, such singular solutions can give some hints toward a resolution of the information loss paradox.

We have observed that there are two possible limits around the singularity, the weak coupling limit and the strong coupling limit. In the former case, one can regard the singularity as a firewall. Due to this firewall-like surface, there is no genuine interior of the black hole and the overall spacetime is causally connected, hence there may be no information loss. In the latter case, due to the strong coupling, energy should be emitted rapidly and the singularity will disappear. Even if the singularity is unstable, if there exists a star-like interior and if it prevents the formation of such a singular structure for some time, then it can mimic a black hole and can help to explain the information loss problem.

Our model is quite naturally realizable within the context of string theory. On the other hand, there are some possible criticisms:
\begin{itemize}
\item[--] Why is there a non-trivial $b$? In other words, why does $b$ have a non-zero value? Is there any fundamental reason to make $b$ non-vanishing?
\item[--] Is the star-like solution realizable in real stars? Can we regard the model as sufficiently realistic and not too artificial?
\end{itemize}

At this moment, we cannot answer all questions about the generality and the realism of the solutions. However, if such solutions are allowed in string theory, they will show interesting behavior in terms of the information loss problem: if our analysis is right, then these solutions probably do not suffer from information loss. Therefore, this may shed some light on the problem. Even though it cannot be a generic resolution, we need to take seriously the treatment of the behavior of such singular solutions. The extension toward a more generic context of string-inspired models is also an interesting topic. These issues are postponed for future works.

\section*{Acknowledgments}
The authors would like to thank Jeong-Hyuck Park, Stephen Angus, and Alexey Golovnev for stimulating discussions and useful comments. DY was supported by Pusan National University Research Grant, 2019.

\section*{Appendix: Comments on tensions}

As we consider the thin-shell approximation, there must be contributions not only from additional matter fields but also from the dilaton field as well as the Kalb-Ramond field. The question is whether these two fields give any singular contribution or not. This appendix is devoted to confirm that there is no such a singular effect.

From \cite{Ko:2016dxa}, we bring the Einstein equations that have non-trivial differentiation terms as follows:
\begin{eqnarray}
8\pi T_{tt} &=& \frac{1}{2} A'A \frac{d}{dr} \ln\left(\frac{A'}{A}C\right) + \phi' A^{2} \frac{d}{dr} \ln (\phi'C) - \frac{1}{2} h^{2} \frac{A^{2}}{C^{2}} e^{-4\phi},\\
8\pi T_{rr} &=& \frac{1}{2} \frac{A'}{A} \frac{d}{dr} \ln\left(\frac{A'}{A}C\right) - \frac{1}{2} \frac{A'^{2}}{A^{2}} - \frac{1}{\sqrt{C}} \frac{d}{dr} \left(\frac{C'}{\sqrt{C}}\right) - 2\phi'^{2} - \phi' \frac{d}{dr} \ln \left( \phi'C \right) - \frac{1}{2} A^{2}\frac{B^{2}}{C^{2}} e^{-4\phi} ,\\
8\pi T_{\theta\theta} &=& 1 - \frac{1}{2}C'' + \frac{d}{dr} \left( \frac{1}{2} \frac{A'}{A}C - \phi'C \right) - \frac{1}{2} \left( A^{2}B^{2} - h^{2} \right) \frac{e^{-4\phi}}{C},
\end{eqnarray}
where $T_{\mu\nu}$ is the energy-momentum tensor contributions from an additional matter fields (that mimic the star interior) and the metric form is Eq.~(\ref{eq:metform}). In the thin-shell limit, the matter contributions should be expanded such as
\begin{eqnarray}
T_{\mu\nu} = S_{\mu\nu} \delta(r - \bar{r}) + (\mathrm{regular\;terms}),
\end{eqnarray}
if the shell is located at $r = \bar{r}$. We ask whether there is any singular term which is proportional to $\delta(r-\bar{r})$.

In the thin-shell limit, if we denote the thickness of the shell is approximately $\Delta r$, one can approximate
\begin{eqnarray}
\frac{1}{\Delta r} &\rightarrow& \delta(r - \bar{r}).
\end{eqnarray}
Since there is no gradient of $h$, there is no contribution to $S_{\mu\nu}$ from the Kalb-Ramond field. Regarding the dilaton field,
\begin{eqnarray}
\phi'' &\simeq& \frac{\Delta \phi}{\Delta r^{2}} \rightarrow \frac{\Delta \phi}{\Delta r} \delta(r-\bar{r}),\\
\phi' &\simeq& \frac{\Delta \phi}{\Delta r},
\end{eqnarray}
where $\Delta \phi \simeq \phi_{+} - \phi_{-}$. Hence, if $\Delta r \simeq \Delta \phi \ll \bar{r}$, i.e., the field value difference and the thickness of the shell is sufficiently smaller than the size of the shell, we can denote $\Delta \phi/\Delta r$ as a regular function of $r$. In this limit, the thin-shell approximation is not singular and we can trust the approximation. As long as there is no such a singular behavior from the dilaton or the Kalb-Ramond field (and, of course, if the contributions from the dilaton and the Kalb-Ramond fields are maintained sufficiently small), it is reasonable to assume that the dominant contribution to the tension or the pressure is originated from the additional matter fields that mimic the star interior.


\newpage


\begin{thebibliography}{200}

\bibitem{Hawking:1974sw} 
  S.~W.~Hawking,
  Commun.\ Math.\ Phys.\  {\bf 43}, 199 (1975)
  Erratum: [Commun.\ Math.\ Phys.\  {\bf 46}, 206 (1976)].

\bibitem{Hawking:1976ra} 
  S.~W.~Hawking,
  Phys.\ Rev.\ D {\bf 14}, 2460 (1976).

\bibitem{Maldacena:1997re} 
  J.~M.~Maldacena,
  Int.\ J.\ Theor.\ Phys.\  {\bf 38}, 1113 (1999)
  [Adv.\ Theor.\ Math.\ Phys.\  {\bf 2}, 231 (1998)]
  [hep-th/9711200].

\bibitem{Banks:1983by} 
  T.~Banks, L.~Susskind and M.~E.~Peskin,
  Nucl.\ Phys.\ B {\bf 244}, 125 (1984);\\
  W.~G.~Unruh and R.~M.~Wald,
  Phys.\ Rev.\ D {\bf 52}, 2176 (1995)
  [hep-th/9503024].

\bibitem{Susskind:1993if} 
  L.~Susskind, L.~Thorlacius and J.~Uglum,
  Phys.\ Rev.\ D {\bf 48}, 3743 (1993)
  [hep-th/9306069].

\bibitem{Yeom:2008qw} 
  D.~Yeom and H.~Zoe,
  Phys.\ Rev.\ D {\bf 78}, 104008 (2008)
  [arXiv:0802.1625 [gr-qc]];\\
  S.~E.~Hong, D.~Hwang, E.~D.~Stewart and D.~Yeom,
  Class.\ Quant.\ Grav.\  {\bf 27}, 045014 (2010)
  [arXiv:0808.1709 [gr-qc]];\\
  D.~Yeom and H.~Zoe,
  Int.\ J.\ Mod.\ Phys.\ A {\bf 26}, 3287 (2011)
  [arXiv:0907.0677 [hep-th]].

\bibitem{Almheiri:2012rt} 
  A.~Almheiri, D.~Marolf, J.~Polchinski and J.~Sully,
  JHEP {\bf 1302}, 062 (2013)
  [arXiv:1207.3123 [hep-th]];\\
  A.~Almheiri, D.~Marolf, J.~Polchinski, D.~Stanford and J.~Sully,
  JHEP {\bf 1309}, 018 (2013)
  [arXiv:1304.6483 [hep-th]].

\bibitem{Hwang:2012nn} 
  D.~Hwang, B.~H.~Lee and D.~Yeom,
  JCAP {\bf 1301}, 005 (2013)
  [arXiv:1210.6733 [gr-qc]];\\
  W.~Kim, B.~H.~Lee and D.~Yeom,
  JHEP {\bf 1305}, 060 (2013)
  [arXiv:1301.5138 [gr-qc]];\\
  P.~Chen, Y.~C.~Ong, D.~N.~Page, M.~Sasaki and D.~Yeom,
  Phys.\ Rev.\ Lett.\  {\bf 116}, no. 16, 161304 (2016)
  [arXiv:1511.05695 [hep-th]].

\bibitem{Chen:2014jwq} 
  P.~Chen, Y.~C.~Ong and D.~Yeom,
  Phys.\ Rept.\  {\bf 603}, 1 (2015)
  [arXiv:1412.8366 [gr-qc]].

\bibitem{Hwang:2016otg} 
  J.~Hwang, D.~S.~Lee, D.~Nho, J.~Oh, H.~Park, D.~Yeom and H.~Zoe,
  Class.\ Quant.\ Grav.\  {\bf 34}, no. 14, 145004 (2017)
  [arXiv:1608.03391 [hep-th]];\\
  J.~Hwang, H.~Park, D.~Yeom and H.~Zoe,
  J.\ Korean Phys.\ Soc.\  {\bf 73}, no. 10, 1420 (2018)
  [arXiv:1712.00347 [hep-th]].

\bibitem{Hawking:2005kf} 
  S.~W.~Hawking,
  Phys.\ Rev.\ D {\bf 72}, 084013 (2005)
  [hep-th/0507171].

\bibitem{Sasaki:2014spa} 
  M.~Sasaki and D.~Yeom,
  JHEP {\bf 1412}, 155 (2014)
  [arXiv:1404.1565 [hep-th]];\\
  B.~H.~Lee, W.~Lee and D.~Yeom,
  Phys.\ Rev.\ D {\bf 92}, no. 2, 024027 (2015)
  [arXiv:1502.07471 [hep-th]];\\
  P.~Chen, G.~Domenech, M.~Sasaki and D.~Yeom,
  JCAP {\bf 1604}, no. 04, 013 (2016)
  [arXiv:1512.00565 [hep-th]].
  
\bibitem{Chen:2017suz} 
  P.~Chen, G.~Domenech, M.~Sasaki and D.~Yeom,
  JHEP {\bf 1707}, 134 (2017)
  [arXiv:1704.04020 [gr-qc]];\\
  P.~Chen, M.~Sasaki and D.~Yeom,
  arXiv:1806.03766 [hep-th].
  
\bibitem{Maldacena:2013xja} 
  J.~Maldacena and L.~Susskind,
  Fortsch.\ Phys.\  {\bf 61}, 781 (2013)
  [arXiv:1306.0533 [hep-th]].

\bibitem{Chen:2016nvj} 
  P.~Chen, C.~H.~Wu and D.~Yeom,
  JCAP {\bf 1706}, no. 06, 040 (2017)
  [arXiv:1608.08695 [hep-th]];\\
  S.~Kang and D.~Yeom,
  Phys.\ Rev.\ D {\bf 97}, no. 12, 124031 (2018)
  [arXiv:1703.07746 [gr-qc]].


\bibitem{Burgess:1994kq} 
  C.~P.~Burgess, R.~C.~Myers and F.~Quevedo,
  Nucl.\ Phys.\ B {\bf 442}, 75 (1995)
  [hep-th/9410142].

\bibitem{Siegel:1993fxa} 
  W.~Siegel,   
  Phys.\ Rev.\ D {\bf 47},  5453 (1993)
  [arXiv:9302036 [hep-th]];\\
  W.~Siegel, 
  Phys.\ Rev.\ D {\bf 48}, 2826 (1993)
  [arXiv:9305073 [hep-th]];\\
  C.~Hull and B.~Zwiebach, 
  JHEP {\bf 0909}, 099 (2009)
  [arXiv:0904.4664 [hep-th]];\\
  C.~Hull and B.~Zwiebach, 
  JHEP {\bf 0909}, 090 (2009)
  [arXiv:0908.1792 [hep-th]];\\
  O.~Hohm, C.~Hull and B.~Zwiebach, 
  JHEP {\bf 1007}, 016 (2010)
  [arXiv:1003.5027 [hep-th]];\\
  O.~Hohm, C.~Hull and B.~Zwiebach, 
  JHEP {\bf 1008},  008 (2010)
  [arXiv:1006.4823 [hep-th]].

\bibitem{Ko:2016dxa} 
  S.~M.~Ko, J.~H.~Park and M.~Suh,
  JCAP {\bf 1706}, no. 06, 002 (2017)
  [arXiv:1606.09307 [hep-th]];\\
  S.~Angus, K.~Cho and J.~H.~Park,
  Eur.\ Phys.\ J.\ C {\bf 78}, no. 6, 500 (2018)
  [arXiv:1804.00964 [hep-th]].
  
\bibitem{Mathur:2005a} 
  D.~Mathur,
  Fortsch.Phys. 53 (2005) 793-827
  [arXiv:hep-th/0502050 [hep-th]].
    
\bibitem{Davies:1976ei} 
  P.~C.~W.~Davies, S.~A.~Fulling and W.~G.~Unruh,
  Phys.\ Rev.\ D {\bf 13}, 2720 (1976).

\bibitem{Israel:1966rt} 
  W.~Israel,
  Nuovo Cim.\ B {\bf 44S10}, 1 (1966)
  [Nuovo Cim.\ B {\bf 44}, 1 (1966)]
  Erratum: [Nuovo Cim.\ B {\bf 48}, 463 (1967)].

\bibitem{Garcia:2011aa} 
  N.~M.~Garcia, F.~S.~N.~Lobo and M.~Visser,
  Phys.\ Rev.\ D {\bf 86}, 044026 (2012)
  [arXiv:1112.2057 [gr-qc]].

\bibitem{Hansen:2009kn} 
  J.~Hansen, D.~Hwang and D.~Yeom,
  JHEP {\bf 0911}, 016 (2009)
  [arXiv:0908.0283 [gr-qc]];\\
  D.~Hwang and D.~Yeom,
  Class.\ Quant.\ Grav.\  {\bf 28}, 155003 (2011)
  [arXiv:1010.3834 [gr-qc]].
  


\end{thebibliography}
\end{document}